\documentclass[
twocolumn,
]{ceurart}

\usepackage{listings}

\definecolor{lightBlue}{RGB}{0, 153, 255}
\definecolor{lightRed}{RGB}{204, 0, 255}

\lstdefinelanguage{Manchester}
{
    sensitive = true,
    keywords = [1]{Class, EquivalentTo, SubClassOf, DisjointUnionOf, Annotations, DisjointWith},
    morekeywords = [2]{and, or, not},
    morekeywords = [3]{some, value, only, min, max, exactly},
    keywordstyle=[2]\textbf,
    keywordstyle=[2]\color{lightBlue}\textbf,
    keywordstyle=[3]\color{lightRed}\textbf,
    morestring=[b]"
}

\usepackage{hyperref}

\begin{document}

\copyrightyear{2021}
\copyrightclause{Copyright for this paper by its authors.
  Use permitted under Creative Commons License Attribution 4.0
  International (CC BY 4.0).}

\conference{Workshop on Personal Knowledge Graphs Co-located with the 3rd Automatic Knowledge Base Construction Conference (AKBC'21)}

\title{Personal Health Knowledge Graph for Clinically Relevant Diet Recommendations}

\author[1]{Oshani Seneviratne}[%
orcid=0000-0001-8518-917X,
email=senevo@rpi.edu,
]
\address[1]{Rensselaer Polytechnic Institute,
  110 8th Street, Troy NY 12180, USA}

\author[1]{Jonathan Harris}[%
email=harrij15@rpi.edu,
]

\author[2]{Ching-Hua Chen}[%
orcid=0000-0002-1020-0861,
email=chinghua@us.ibm.com,
]
\address[2]{
Center for Computational Health, IBM Research, Yorktown Heights, NY, USA}

\author[1]{Deborah L. McGuinness}[%
orcid=0000-0001-7037-4567,
email=dlm@cs.rpi.edu,
]

\begin{abstract}
We propose a knowledge model for capturing dietary preferences and personal context to provide personalized dietary recommendations. 
We develop a knowledge model called the Personal Health Ontology, which is grounded in semantic technologies, and represents a patient's combined medical information, social determinants of health, and observations of daily living elicited from interviews with diabetic patients. 
We then generate a personal health knowledge graph that captures temporal patterns from synthetic food logs, annotated with concepts from the Personal Health Ontology.
We further discuss how lifestyle guidelines grounded in semantic technologies can be reasoned with the generated personal health knowledge graph to provide appropriate dietary recommendations that satisfy the user's medical and other lifestyle needs.
\end{abstract}


\maketitle

\section{Introduction}

\begin{figure*}[!htbp]
    \centering
    \includegraphics[width=\textwidth]{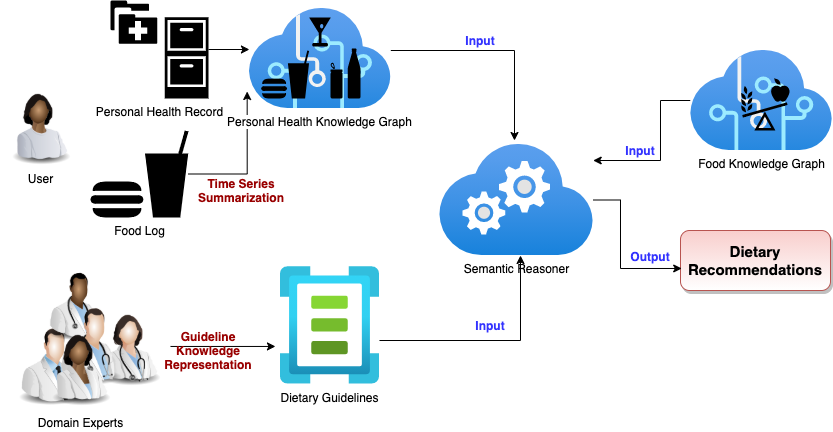}
    \caption{Overview of the System Generating Dietary Recommendations}
    \label{fig:dietary_recommendation_overview}
\end{figure*}

Chronic illnesses typically involve multiple healthcare specialists over an extended period. 
Relying on Electronic Health Records (EHR) alone to understand the health status of a person with a chronic illness could result in an incomplete picture of patient profiles~\cite{madden2016missing}.
Such an incomplete picture is not only due to disconnected EHR systems being used by different healthcare providers, but also due to incomplete data from Observations of Daily Living (ODL)~\cite{backonja2012observations} and Social Determinants of Health (SDoH)~\cite{roux2015social}. 
Many chronic illnesses, such as Type 2 Diabetes (T2D), impose a significant burden on patients to self-manage their disease, requiring health behavior change, health status monitoring with robust ODL mechanisms, medication adherence, and patient self-management behaviors that may require comprehensive education. 
Most clinicians focus their efforts on medical care while expecting patients to take on the responsibility of lifestyle modifications, including healthy eating.
Therefore, using nutritional behavior as a focus, we demonstrate how semantic modeling of nutritional guidelines and semantic reasoners can utilize the Personal Health Knowledge Graph (PHKG) to provide personalized insights for T2D self-management consistent with clinical guidelines.
Our PHKG is annotated using the Personal Health Ontology (PHO) developed for capturing lifestyle behaviors related to food consumption.
Furthermore, using an application that captures summarized food logs annotated with concepts from 
our FoodKG~\cite{haussmann2019foodkg} and 
our PHO-annotated PHKG, we run a semantic reasoner to generate clinically relevant dietary recommendations for the user as demonstrated in Figure~\ref{fig:dietary_recommendation_overview}.

\section{Related Work}

A Personal Knowledge Graph (PKG) as defined by \citeauthor{balog2019personal} is viewed as a consolidated resource that stores unique entities, which may additionally be linked to external sources and contains entities that do not exist in any other knowledge graphs~\cite{balog2019personal}. 
The research proposal set out by \citeauthor{balog2019personal} for PKGs further outlines several key differences between personalized KGs and personal KGs. Personalized KGs do not go beyond a general-purpose KG, but rather customize it in some way to match a user’s need or profile. 
Personal KGs, on the other hand, contain a set of entities with links to other KGs, capturing many aspects centered around a user. 
Such personalization is essential when providing clinically relevant dietary recommendations for an individual, which is our work's focus.
For example, \citeauthor{balog2019personal} highlights that PKG relations may often be short-lived (e.g. ``what I plan on cooking tonight'').
In our PHO, we capture the semantics that can potentially answer such questions.

Extending the PKGs to the health domain, \citeauthor{gyrard2018personalized} define Personal \emph{Health} Knowledge Graph to include information about people, including their medical history, demographics, social information, and preferences~\cite{gyrard2018personalized}. 
The PHKG is then a subgraph of a larger KG that aggregates knowledge from heterogeneous sources, including medical datasets and IoT devices containing only information relevant to the user. 
However, given that broad definition, it is often impossible to determine what would be included and excluded from the graph, especially in a dietary recommendation setting.

The Healthy Lifestyle Support (HeLiS) Ontology~\cite{dragoni2018helis} integrates food and activity for personalized health monitoring scenarios and has been utilized in a mobile application that can be used by patients with chronic diseases such as T2D.
However, the HeLiS ontology modeling primarily focuses on the specific foods and the activities the individuals must consume, not on the contextual elements that could be used to personalize any recommendations given to the patient.

The Ontology for Nutritional Studies (ONS) integrates the terms related to food description, medical science, genetics, genomics data, and nutritional science methods for diet and health research~\cite{vitali2018ons}. 
The ONS was developed to harmonize biochemical, genetic, clinical, and nutritional concepts typically found in intervention and observational studies in human nutrition that can assist nutrition researchers by selecting the appropriate terms from a wide range of existing ontologies and creating the relevant missing key concepts for the field.
While this ontology contains a lot of information about human nutrition, due to the focus of observational studies, it is not straightforward to use it for dietary recommendations.
In addition, the Bionutrition Ontology (BNO) represents a controlled vocabulary of nutritional terms~\cite{yang2019ontology}.
However, the BNO lacks proper annotation of terms or definitions of properties and lacks orthogonality (i.e., no terms are imported or refer to external ontologies).
Many of these limitations were addressed in the design of the PHO described in Section~\ref{sec:pho}.

\section{Personal Health Ontology (PHO)}
\label{sec:pho}

Based on a set of interviews conducted with 21 people who declared themselves to be within five years of being diagnosed with T2D, we developed the PHO to characterize the dietary behaviors of people with T2D,
with the expectation that the PHO would be extended to other aspects of personal health in the future. 

First, we extracted answers from the interview transcripts and categorized them into concepts that capture the nutritional questions and concerns of people with diabetes. 
Special care was taken not to include general knowledge questions about food \emph{(``What is Tahini made of?'')}, as that could be answered by our FoodKG~\cite{haussmann2019foodkg} directly without any personalization.
The categories include:
\begin{enumerate}
    \item \textbf{Likes} \emph{(I prefer spicy food. Suggest a good breakfast with chilies.)} 
    \item \textbf{Dislikes} \emph{(I do not like peanuts. Suggest a snack without peanuts.)}
    \item \textbf{Nutrient-focused} \emph{(Can you suggest Indian recipes with fewer calories [due to my T2D condition]?)}
    \item \textbf{Substitution} \emph{(What should I eat instead [due to my T2D condition]?)}
    \item \textbf{Restaurant Knowledge} \emph{(To keep to my dietary regimen, what restaurants should I go to?)} 
    \item \textbf{Cooking Methodology} \emph{(What dishes could I prepare with the appliances I have in my kitchen?)}
    \item \textbf{Financial} \emph{(Are there cheaper menu items that I can eat?)}
    \item \textbf{Medical Condition} \emph{(Can you suggest recipes that would limit my vitamin K intake?)}
\end{enumerate}

Second, we put a particular emphasis on the following aspects that were underlying themes across many of the questions and answers of the participants.

\noindent\textbf{Context: } 
It might be that little more is known about an entity than its type and the relation to the user. For example, the likes and dislikes of the user can be ascertained by looking at the foods that co-occur in the food log. The favorite food type can be determined by looking at the frequency of the meal.
The PHKG should not contain all the possible entity attributes but only those important to the user in a given context.
To encode such types of context, we primarily used the concepts and relationships from the Provenance (prov) ontology (\url{https://www.w3.org/TR/prov-o}).
 
\noindent\textbf{Time: } 
Well-established and relatively stable relationships over time are essential criteria to be considered for inclusion in the PHKG.
Therefore, we defined concepts such as \texttt{ConsistentPattern} to capture temporal patterns to identify the user's behaviors to either suggest similar foods (if they are healthy options) or alternate foods (if they are not appropriate given the user's specific health condition).

\noindent\textbf{Likelihood Estimates: } 
In some cases, it would be desirable to include the likelihood of a personal recommendation or the temporal patterns mined from their food logs to someone.
For those purposes, we utilize several concepts from community-defined ontologies. Examples include the \texttt{coefficientOfVariation} concept from the Statistics Ontology (stato) (\url{http://stato-ontology.org}), and \texttt{frequency} from the Semantic-science Integrated Ontology (sio) (\url{https://bioportal.bioontology.org/ontologies/SIO}).

\section{Personal Health Knowledge Graph (PHKG) Generation}
\label{sec:phkg}

Utilizing an extended Time Series Summarization (TSS) framework~\cite{harris2021framework} on synthetic food logs, we generated behavioral insights annotated with the PHO to generate the PHKG focusing on dietary behaviors of users.
The TSS framework was previously designed to automatically generate natural language summaries of temporal personal health data utilizing advanced temporal data-mining techniques, such as frequent pattern mining and categorical clustering methods, to surface comprehensible and meaningful explanations to a non-expert user~\cite{harris2021framework}. We extended this framework to generate RDF triples based on temporal patterns found within the data. The patterns discovered are related to the constraints mentioned by T2D patients within the user study mentioned above. 
Instead of the natural language summaries that would be output from the TSS, we utilize the PHO concepts to capture the temporal patterns in the PHKG as RDF triples.
The food log was gathered from synthetic data spanning five weeks at the meal-level granularity using the MyFitnessPal schema. For each meal, we have the nutrient information of the food consumed (i.e., nutrient consumption such as calorie intake) and the names of the foods consumed. 
Since our goal is to recommend clinically relevant diet recommendations, we focus on temporal patterns on nutrient intake, such as the recommendation to maintain a consistent carbohydrate intake, such as the following temporal summary.

\lstset{language=Manchester, basicstyle=\ttfamily\fontsize{9}{10}\selectfont, columns=fullflexible, xleftmargin=5mm, framexleftmargin=5mm, breaklines=true, numbers=left, numberstyle=\ttfamily\fontsize{9}{10}\selectfont, numbersep=5pt, showstringspaces=false, frame=lines, caption={RDF representation for ``\emph{This past full week (starting from Sep 23 and ending on Sep 30, you have kept your carbohydrate intake relatively fixed (or consistent).}''}, label=lst:carb-intake}
\begin{lstlisting}
:user a prov:Person.
:user sio:hasAttribute :ConsistentCarbohydrateIntake.
:ConsistentCarbohydrateIntake a stato:coefficientOfVariation;
     sio:hasValue "0.99"^^xsd:float;
     prov:startedAtTime "2021-09-23T00:00:00-00:00"^^xsd:dateTime;
     prov:endedAtTime "2021-09-30T00:00:00-00:00"^^xsd:dateTime.
\end{lstlisting}

A user whose dietary preference is to maintain a low-carb, high-fat diet would be interested in a summary that specifies how frequently they have been consistent with a low-carb, high-fat diet. 
The concepts \texttt{LowCarbDiet}, and \texttt{highFatDiet} are defined in the PHO to comprise of a diet either high or low of the corresponding macro-nutrient as advised by a subject matter expert. The extended-TSS uses these threshold values to determine if the temporal data pertaining to the diet is either a \texttt{LowCarbDiet} and/or \texttt{highFatDiet}, and labels it as such.
\lstset{language=Manchester, basicstyle=\ttfamily\fontsize{9}{10}\selectfont, columns=fullflexible, xleftmargin=5mm, framexleftmargin=5mm, breaklines=true, numbers=left, numberstyle=\ttfamily\fontsize{9}{10}\selectfont, numbersep=5pt, showstringspaces=false, frame=lines, caption={RDF representation for ``\emph{You have been maintaining a low-carb, high-fat diet.}''}, label=lst:low-carb-high-fat}
\begin{lstlisting}
:user sio:hasAttribute :LowCarbDiet, :highFatDiet.
:LowCarbDiet sio:frequency "1.0"^^xsd:float.
:HighFatDiet sio:frequency "1.0"^^xsd:float.
\end{lstlisting}

We can also go deeper by finding a relationship between low-carb and high-fat consumption on a day-to-day basis using a relationship that encapsulates the consumption of the two types of macronutrients that would be captured in the \texttt{:LowCarbHighFatNutrientIntakeGoal}.
A boolean value could indicate whether the user usually tends to exhibit either kind of consumption, as shown in Listing~\ref{lst:low-carb-high-fat-consumption-on-sunday}.  

\lstset{language=Manchester, basicstyle=\ttfamily\fontsize{9}{10}\selectfont, columns=fullflexible, xleftmargin=5mm, framexleftmargin=5mm, breaklines=true, numbers=left, numberstyle=\ttfamily\fontsize{9}{10}\selectfont, numbersep=5pt, showstringspaces=false, frame=lines, caption={RDF representation for ``\emph{You have been maintaining a low-carb, high-fat diet.}''}, label=lst:low-carb-high-fat-consumption-on-sunday}
\begin{lstlisting}
:user sio:hasAttribute :LowCarbHighFatNutrientIntakeGoal.
:LowCarbHighFatNutrientIntakeGoal 
    sio:hasParticipant :LowCarbDiet, :HighFatDiet;
    sio:hasValue "true"^^xsd:boolean.
\end{lstlisting}

Our PHKG comprises patterns such as the ones described above captured as RDF triples, with a specific focus on foods consumed.

\section{Dietary Guidelines Modeling}
\label{sec:dietary-guidelines}

For T2D treatment and management, the American Diabetes Association (ADA)~\cite{american20201} has generated a set of guidelines that includes the necessary guidance of dietary intake for people with diabetes or pre-diabetes.
As such, the ADA clinical practice guideline contains several \emph{chapters}, such as ``Classification and Diagnosis of Diabetes"~\cite{american20202} that includes information on characterizing diabetes, ``Prevention or Delay of Type 2 Diabetes"~\cite{american20203} that includes lifestyle interventions, among others, to guide pre-diabetic individuals, and ``Pharmacologic Approaches to Glycemic Treatment"~\cite{american20209} that contains information for managing T2D with medications.

In this section, we describe the semantic modeling of two selected ADA guideline recommendations focused on dietary recommendations.
Each ADA guideline recommendation was captured in two parts: (1) rule indicates the necessary and sufficient conditions for a guideline to be in a compliant state, and (2) directive indicates what action to take if the rule was evaluated to be non-compliant. 

\subsection{Example Guidelines}
\label{sec:example-guidelines}
\noindent\textbf{Guideline 1:}
\noindent\textbf{Guideline 1:}

\begin{quote}
    \textit{``For pre-diabetic and diabetic individuals, diet low in total fat but relatively high in carbohydrates should be replaced with Mediterranean diet.''}
\end{quote}

The semantic encoding of this guideline contains the rules (conditions) for when this recommendation should be valid can be identified. 
These include (i) the person has T2D or pre-diabetes (since we are applying the rule from ADA), (ii) the person is consuming a diet, and (iii) the diet is classified as a \texttt{HighCarbDiet} and \texttt{LowFatDiet}. 
In addition, the recommended action specifies that the diet should be replaced with a Mediterranean diet.

\lstset{language=Manchester, basicstyle=\ttfamily\fontsize{9}{10}\selectfont, columns=fullflexible, xleftmargin=5mm, framexleftmargin=5mm, numbers=left, stepnumber=1, breaklines=true, breakatwhitespace=false, numberstyle=\ttfamily\fontsize{9}{10}\selectfont, numbersep=5pt, tabsize=2, frame=lines, caption={OWL expression of Guideline 1.}, label=lst-dietary-guideline-rule1-modeling}
\begin{lstlisting}
Class: Diabetic
  EquivalentTo: 
    DietaryAssessment and
    prov:wasAssociatedWith some prov:Person and
    prov:wasAssociatedWith doid:Diabetes or doid:PreDiabetes
  SubClassOf: 
    DiabeticStatusAssessment
Class: DiabeticLowFatHigCarbDietConsumption
  EquivalentTo: 
    Diabetic and
    sio:hasAttribute some 
        (ConsistentPattern
            and (sio:hasAttribute only 
                (HighCarbDiet and LowFatDiet)))
  SubClassOf:
    DietConsumption

Class: MediterraneanDietDirective
  EquivalentTo: 
    prov:wasAssociatedWith some 
      (sio:hasAttribute some DiabeticLowFatHigCarbDietConsumption and MediterraneanDietRecommendation)
  SubClassOf: 
    Directive

Class: MediterraneanDietRecommendation
    EquivalentTo: 
        Constraint only 
            "{tag: 'Mediterranean'}"
    Annotations: 
    rdfs:label "For pre-diabetic and diabetic individuals diet low in total fat but relatively high in carbohydrates should be replaced with Mediterranean diet."
  SubClassOf: 
    Recommendation
\end{lstlisting}

\noindent\textbf{Guideline 2:}

\begin{quote}
    \textit{``For individuals whose daily insulin dosing is fixed, a consistent pattern of carbohydrate intake with respect to time and amount may be recommended to improve glycemic control and reduce the risk of hypoglycemia.''}
\end{quote}

In the above rule, if a T2D patient is undergoing insulin therapy, they must have a consistent carbohydrate intake. The rule captures insulin intake and food consumption over a given period and determines whether the carbohydrate intake has been consistent. 
As demonstrated by the rule's OWL encoding in Listing~\ref{lst-dietary-guideline-rule2-modeling}, there is a temporal pattern descriptor (i.e., \texttt{ConsistentCarbPattern}) that is associated with the consistent consumption of an amount of carbohydrates during a particular meal
(e.g., breakfast) as determined by the TSS. 
There are also specific personal characteristics of a user (i.e., \texttt{Diabetes} status and \texttt{FixedInsulinDosage}), all of which are modeled in the PHO and captured in the PHKG.
The \texttt{recommendation} provides the required range as a personalized guideline constraint, which would be used in the downstream application in providing the relevant dietary recommendation.

\lstset{language=Manchester, basicstyle=\ttfamily\fontsize{9}{10}\selectfont, columns=fullflexible, xleftmargin=5mm, framexleftmargin=5mm, numbers=left, stepnumber=1, breaklines=true, breakatwhitespace=false, numberstyle=\ttfamily\fontsize{9}{10}\selectfont, numbersep=5pt, tabsize=2, frame=lines, caption={OWL expression of Guideline 2.}, label=lst-dietary-guideline-rule2-modeling}
\begin{lstlisting}
Class: FixedInsulinDosage
  EquivalentTo: 
    sio:hasAttribute dron:Insulin, 
    FixedMedicationDosage

Class: ConsistentCarbPattern
  EquivalentTo: 
    ConsistentPattern and
    (sio:hasAttribute some food:Carbohydrates)

Class: ConsistentCarbDietDirective
  EquivalentTo: 
    Diabetic and
    (sio:hasAttribute some FixedInsulinDosage) and   
    (sio:hasAttribute some ConsistentCarbPattern) and
  prov:wasAssociatedWith some 
      (sio:hasAttribute some  ConsistentCarbRecommendation)

Class: ConsistentCarbRecommendation
    EquivalentTo: 
        Constraint only 
            "{'carbohydrate' : 
                {'unit': 'g',
                 'meal' : 
                   {'type': 'range',
                   'lower' : '30',
                   'upper': '45'},
                'daily total' : '150'}}" .
  Annotations: 
    rdfs:label "For individuals whose daily insulin dosing is fixed, a consistent pattern of carbohydrate intake with respect to time and amount may be recommended to improve glycemic control and reduce the risk of hypoglycemia."
\end{lstlisting}

\subsection{Semantic Reasoner in Action}

Using the guideline rules modeled in Section~\ref{sec:example-guidelines} on the PHKG generated in Section~\ref{sec:phkg}, we can generate directives that provide clinically relevant dietary recommendations.
Specifically, the semantic reasoner would assert a specific subclass of the \texttt{Directive} (e.g., \texttt{MediterraneanDietDirective} or \texttt{ConsistentCarbDietDirective}).
These asserted directives would be associated with a certain \texttt{Recommendation} (e.g., \texttt{MediterraneanDietRecommendation} and \texttt{ConsistentCarbRecommendation}), that would inform a downstream application how to provide an appropriate recommendation.
We provide several questions that could be translated into SPARQL and evaluated with the insights generated from the semantic reasoner in Section~\ref{sec:competency-questions}.

\section{Evaluation with Competency Questions}
\label{sec:competency-questions}

We utilize a competency question-based evaluation methodology~\cite{ren2014towards} for evaluating the PHKG generated. Below, we describe three types of questions and how the answers can be derived from the PHKG.
All of these questions have a corresponding SPARQL representation. In a similar vein, as long as a natural language question can be represented using the terms available in the PHO, an answer can be derived.

\subsection{Performance}

These types of questions take the following form:
\begin{enumerate}
    \item \textbf{Progress:} \emph{(``How have  I been doing (improving, getting worse, maintaining) over the past day/week?)''}
    \item \textbf{Consistency:} \emph{(``Have I been consistent in my carbohydrate intake?'')}
    \item \textbf{Compliance:} \emph{(``Have I been following a Mediterranean diet?'')}
\end{enumerate}

To answer these types of questions related to the user's performance, we need to focus on how well their dietary intake matches up with the guidelines within a certain period. For example, the user may be struggling to maintain a consistent carbohydrate intake over the past week because they have been missing breakfast. In such a scenario, a question such as \#1 above can be answered by discovering patterns within relevant temporal personal health data and deciding which of these may be the most important to surface to the user.

\subsection{Behavioral Recommendations}

\begin{enumerate}
    \item \textbf{Improve Diet:} \emph{(``How can I improve my diet strategy (considering personal preferences and context)?'')}
    \item \textbf{Improve Performance:} \emph{(``Will my current diet strategy improve my performance?'')}
    \item \textbf{Satisfying Preferences:} \emph{(``Does my current diet strategy meet my preferences?'')}
\end{enumerate}

To successfully answer these questions, we include the user's personal preferences and context in the PHKG. For example, if the user cannot eat breakfast due to their demanding daily schedule, the system has to come up with alternatives to alleviate this problem by recommending different carbohydrate amounts for lunch and dinner or recommend mid-morning snacks.

\subsection{Food Recommendations}

\begin{enumerate}
    \item \textbf{Use Implicit Knowledge:} \emph{(``What should I eat for breakfast?'')}
    \item \textbf{Allergies:} \emph{(``What foods can I eat if I have a dairy allergy?'')}
    \item \textbf{Dislikes:} \emph{(``What can I substitute for almonds?'')}
\end{enumerate}

For question \#1 above, even though the question appears generic, using the PHKG, we augment the question with some implicit knowledge available in the PHKG and expand the question to a form such as \emph{``What should I eat for breakfast [diabetic, prefers spicy food, carbohydrates between 30-45 g, not to exceed 150 g daily total]?''}.
The constraints that go inside the ``[]'' are determined using a semantic reasoner (in our workflow, we used the in-built reasoner in Prot\'{e}g\'{e}~\cite{protege}, and OWLReady2~\cite{lamy2017owlready} for this purpose).
The semantic reasoner evaluates the PHKG against the guidelines to generate the user’s dietary needs and preferences. These will be in the form of personalized guideline constraints that provide input to downstream machine learning tasks such as recipe recommendations. 
In conjunction with the FoodKG~\cite{haussmann2019foodkg} containing over 1 million recipes, ingredients, and nutrients, we can then recommend an appropriate food item either using SPARQL queries or deep learning based personalized food recommendation methods such as pFoodReq~\cite{chen2021personalized}. Both of these approaches would be able to leverage the food preferences, eating habits encoded in the PHKG, and the dietary guidelines or restrictions as appropriate for the user.  
Questions \#2 and \#3 would follow the same process and include the explicit constraints (i.e., allergies and dislikes) stated explicitly.

\section{Conclusion}
There is a need for a system capable of discovering dietary insights hidden in a user's temporal personal health data, such as food logs, and understanding these insights within the user's health context to provide relevant lifestyle recommendations and potentially lead to a personalized dialog. Utilizing the extended time series summarization technique, we generate a Personal Health Knowledge Graph (PHKG) annotated with the Personal Health Ontology for capturing such dietary behaviors. The PHKG, when reasoned using a semantic reasoner, provides performance evaluations against selected lifestyle management guidelines defined by the ADA on appropriate eating habits for the individual.

Our PHKG is different from a regular PKG because it can be readily used in clinically relevant dietary recommendation applications.
The expressivity of the semantic rule representation and the reasoning process enables the recommendation of a food item to a T2D patient. Furthermore, because we have used standards-based ontological terms in the modeling process, the PHKG achieves a high degree of interoperability.

This work could be extended to be used in downstream applications on personalized food recommender systems that utilize task-oriented dialogue such as~\cite{joshi2017personalization}.
Furthermore, even though we have focused on food in this initial modeling of a PHKG, we can adopt a similar approach to extend this work to other domains related to personal health, such as activity, medication intake, and even social determinants of health.
We plan on addressing the many challenges that remain in collecting, managing, integrating, and analyzing the data required to populate, maintain, reason over, explain, and share the PHKG, which makes this an exciting area of research. 

\subsection*{Code and Data Availability}

The Personal Health Ontology, the Extended Time Series Summarization framework, the synthetic Personal Health Knowledge Graph, and the competency questions, along with sample answers, are available at  \url{https://semantics-for-personal-health.github.io}.
\begin{acknowledgments}
This work is supported by IBM Research AI through the AI Horizons Network.
\end{acknowledgments}

\footnotesize{
\bibliography{references}
}

\end{document}